\documentclass{appolb}
\usepackage{mathrsfs}
\usepackage{amssymb,amsmath}
\usepackage{amsfonts}
\usepackage{epsfig}
\def\diag{{\rm diag}}
\def\dmthtw{\Delta m_{32}^2}
\def\dmtwon{\Delta m_{21}^2}
\def\dmthon{\Delta m_{31}^2}

\begin{document}
\pagestyle{plain}

\title{NEUTRINOS, LEPTONIC CP VIOLATION AND THE ORIGIN OF MATTER%
}
\author{F. R. Joaquim
\address{Dipartimento di Fisica \emph{`G.~Galilei'}, Universit\`a di Padova and
INFN-Padova\\ Via Marzolo 8, I-35131 Padua, Italy} } \maketitle

\begin{abstract}
I review some aspects related with the connections between neutrino
physics and the thermal leptogenesis mechanism for the generation of
the cosmological baryon asymmetry of the Universe. A special
attention is devoted to the problem of establishing a bridge between
leptonic $CP$ violation at low and high energies.
\end{abstract}

\section{Introduction}
The problem of why our Universe is dominated by matter has been
object of intense study in the last decades. In spite of satisfying
\emph{\`{a} priori} all the necessary conditions for the
implementation of a successful baryogenesis mechanism, the standard
model (SM) is unable to provide a plausible explanation for the
observed baryon asymmetry of the Universe (BAU). This, and other
longstanding theoretical hints in favor of the existence of physics
beyond the standard model, have been recently supported by one of
the most exciting discoveries of modern particle physics namely,
neutrino oscillations. Apart from presenting us with the challenge
of unraveling the pattern of neutrino masses and mixing, this
discovery may also have a profound impact on our understanding of
the Universe.

Among the various models proposed to explain why neutrinos are
massive (and much lighter than the other known fermions), the seesaw
mechanism \cite{seesaw} has become the most popular due its
simplicity and versatileness. However, and in spite of all the
attempts done in the direction of finding a way to test it, we are
still far from unequivocally select the seesaw mechanism as the one
behind neutrino mass generation. The most promising way of achieving
this goal seems to be the investigation of physical phenomena
capable of constraining the seesaw parameter space. A possible
indirect test relies on the fact that the BAU may have been
generated through the out-of-equilibrium decays of the seesaw heavy
neutrino states via the leptogenesis
mechanism~\cite{Fukugita:1986hr}. Since both the low-energy neutrino
mass and mixing pattern and the value of the BAU depend on the
fundamental seesaw parameters at very high scales, one expects that
for example the amount of $CP$ violation needed to generate a
sufficient BAU is in some sense related with the strength of
low-energy $CP$ violation in the leptonic sector. If this is the
case, future neutrino oscillation experiments will be crucial to
test leptogenesis in the seesaw framework. In this short review I
will analyse some aspects related with the connection between
neutrino physics and the thermal leptogenesis mechanism for the
generation of the BAU.

\section{Seesaw neutrino masses}

The most economical framework over which the seesaw
mechanism~\cite{seesaw} can be realized corresponds to the SM
extended with $n_R$ right-handed neutrino singlets $\nu_{R_j}$. In
the leptonic sector, the SU(2) $\times$ U(1) invariant Yukawa and
mass terms are:
\begin{equation}
-{\cal L}  = \bar{\ell}_{L_i}\widetilde {\phi}\,(
Y_{\nu})_{ij}\,\nu_{R_j}+\bar{\ell}_{L_i}\,\phi\,(Y_{\ell})_{ij}\,
e_{R_j} + \frac{1}{2}\, \nu_{R_i}^{T} C (M_R)_{ij}\,\nu_{R_j} +{\rm
h.c.}\;,\; \label{Lyuk}
\end{equation}
where $\bar{\ell}_{L_i}$ and $e_{R_i}$ stand for the left-handed
lepton doublets and right-handed charged-lepton singlets while
$\phi$ denotes the usual SM lepton Higgs doublet with $\widetilde
{\phi} = i \tau _2 \phi ^\ast$. From now on, I will consider that
the Dirac neutrino Yukawa coupling matrix $Y_{\nu}$ is defined in
the weak basis where the right-handed neutrino mass matrix $M_R$ and
the charged lepton Yukawa couplings $Y_{\ell}$ are diagonal. The
right-handed neutrino mass term is ${\rm SU(2)}\times {\rm U(1)}$
invariant and, as a result, the typical scale of $M_R$ can be much
higher than the electroweak symmetry breaking scale $v=174$~GeV,
thus leading to naturally small left-handed Majorana neutrino
masses. The effective light neutrino mass matrix $\mathcal{M}$ is
given by the well-known seesaw formula
\begin{equation}
\label{meff}%
\mathcal{M}=-v^2\,Y_\nu^{} M_R^{-1}\,Y_\nu^T\;,\;
U^\dag\mathcal{M}\,U^\ast={\rm diag}(m_1,m_2,m_3)\,,
\end{equation}
where I have denoted the light neutrino masses by $m_i$. The unitary
matrix $U$ is the so-called Pontecorvo-Maki-Nakagawa-Sakata (PMNS)
leptonic mixing matrix which can be parameterized by three angles
and three $CP$-violating phases in the following way:
\begin{align}
\label{Uparame}%
U =
U_\delta(\theta_{12},\theta_{13},\theta_{23},\delta)\,P(\alpha,\beta)\;,
\;P(\alpha,\beta)\equiv\diag\,(1,e^{i\alpha/2},e^{i\beta/2})\,.
\end{align}
Here, $\theta_{ij}$ are the leptonic mixing angles and $\delta$ a
Dirac-type $CP$-violating phase. The unitary matrix $U_\delta$ is a
CKM-like mixing matrix and can be written in the standard form as
given for instance in Ref.~\cite{Eidelman:2004wy}. Since the light
neutrinos are predicted to be Majorana particles in the seesaw
scenario, there are two extra Majorana phases $\alpha$ and $\beta$.

The structure of the effective neutrino mass matrix $\mathcal{M}$ is
dictated by $Y_\nu$ and $M_R$. Although the relation between low and
high energy parameters is well established by Eq.~(\ref{meff}), it
is straightforward to see that there is no one-to-one correspondence
between both parameter spaces. Therefore, even if all the entries of
$\mathcal{M}$ are measured, $Y_\nu$ and $M_R$ cannot be
reconstructed. In the most natural scenario where the SM is extended
with three heavy right-handed neutrino singlets, $Y_\nu$ contains 15
parameters and $M_R$ is defined by the three heavy Majorana neutrino
masses $M_i$. In total, there are 18 parameters to be confronted
with the 9 (3 masses+3 mixing angles+3 phases) of $\mathcal{M}$ at
low-energies. This ambiguity inherent to the seesaw mechanism can be
illustrated taking into account that, for fixed $U$, $m_i$ and
$M_i$, the relation
\begin{align}
\mathcal{M}=U\,d_\nu\,U^{T}=-v^2\,Y_\nu^{} d_R^{-1}\,Y_\nu^T\,,
\label{MnuLL}
\end{align}
where $d_\nu=\diag(m_1,m_2,m_3)$ and $d_R=\diag(M_1,M_2,M_3)$, is
always satisfied for~\cite{Casas:2001sr}:
\begin{align}
\label{ynuR}%
Y_\nu=\frac{i}{v}\,U\,\sqrt{d_\nu}\,\mathcal{O}\sqrt{d_R}\;,\;
\mathcal{O}^T \!\mathcal{O}={\bf 1}\,.
\end{align}
Since $\mathcal{O}$ can be any orthogonal complex matrix it is clear
that there is no univocal correspondence between the high and
low-energy neutrino parameters. The meaning of $\mathcal{O}$ can be
suggestively interpreted in terms of the different roles played by
the heavy neutrinos in the seesaw mechanism. In fact, $\mathcal{O}$
can be viewed as a dominance matrix since it gives the weights of
each heavy Majorana neutrino in the determination of the different
light neutrino masses $m_i$~\cite{Smirnov:af}. The fact that
$\mathcal{O}_{ij}^2$ are weights for $m_i$ is quite obvious due to
the orthogonality of $\mathcal{O}$: $m_i  = \sum_j  m_i
\mathcal{O}_{ij}^{\,2}$. On the other hand, the single contribution
$m_i \mathcal{O}_{ij}^{\,2}$ is also given by:
\begin{align}
m_i \mathcal{O}_{ij}^{\,2}= -\frac{(v\,U^\dagger Y_\nu)^2_{ij}}{M_j}
\equiv \frac{X_{ij}}{M_j}\,.
\end{align}
Therefore, once $U$ is settled, each weight $\mathcal{O}_{ij}^2$
just depends on $M_j$ and on its couplings with the left-handed
neutrinos $(Y_\nu)_{kj}$. Consequently, the contribution of each
heavy neutrino to $m_i$ is well defined and expressed by the weight
${\rm Re}(\mathcal{O}_{ij}^2)$. In this sense, one can say that the
heavy Majorana neutrino with mass $M_j$ dominates in $m_i$ if
\begin{align}
\frac{|{\rm Re}\,(X_{ij})|}{M_j} \gg \frac{|{\rm
Re}\,(X_{ik})|}{M_k}\;\;,\;\;k\ne j\,,
\end{align}
which leads to $|{\rm Re}(\mathcal{O}_{ij}^{\,2})| \gg |{\rm
Re}(\mathcal{O}_{ik}^{\,2})|\,$. So, if one of the heavy Majorana
neutrinos gives the dominant contribution to $m_i$, this information
is encoded in the structure of $\mathcal{O}$.

An alternative way to  test the seesaw mechanism\footnote{For
interesting discussions on this subject see
Ref.~\cite{Davidson:2004wi}.} relies on the fact that in the
supersymmetric seesaw new flavor violating effects appear due to the
presence of the Dirac type couplings between the heavy right-handed
neutrinos and the ordinary SM lepton
doublets~\cite{Borzumati:1986qx}. In particular, even in the case
where the supersymmetry breaking mechanism is flavor blind, new
contributions are generated in lepton flavor violating rare decays
like $\ell_i\rightarrow \ell_j\,\gamma$~\cite{LFV}. However, in this
case there is some intrinsic model dependence due to the fact that
neither the high-energy neutrino flavor structure nor the true
mechanism of supersymmetry breaking is known. Nevertheless, future
experimental data will be of extreme value in testing lepton flavor
violation in a wide class of models like those based on grand
unification. In alternative scenarios where, for instance, the
seesaw $mediator$ is a heavy triplet, the results are more
predictive since the amount of flavor violation induced by radiative
corrections is directly related with the low-energy neutrino
data~\cite{Rossi:2002zb}.

\section{Neutrinos and the origin of matter: thermal leptogenesis}

\begin{figure}[t]
\begin{center}
\includegraphics[width=12.cm]{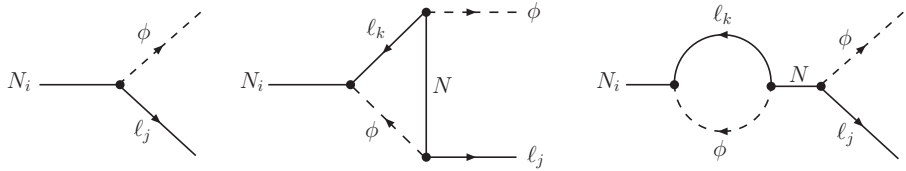}
\caption{Tree-level and one loop contributions to the heavy Majorana
neutrino decays.} \label{Fig1}
\end{center}
\end{figure}

The most recent WMAP results and BBN analysis of the primordial
deuterium abundance imply the following range~\cite{Spergel:2003cb}
\begin{equation}
\eta_{B}=\frac{n_B}{n_\gamma}=(6.1\pm
0.3)\times10^{-10}\,,\label{BAU}
\end{equation}
for the baryon-to-photon ratio of number densities. Although
satisfying the three necessary Sakharov conditions for the
generation of a baryon asymmetry, the SM is not able to justify the
number given in (\ref{BAU}). Among the viable mechanisms to account
for the matter-antimatter asymmetry observed in the Universe,
leptogenesis~\cite{Fukugita:1986hr} has undoubtedly become one of
the most appealing ones. Indeed, its simplicity and close connection
with low-energy neutrino physics render it an attractive and
eventually testable scenario\footnote{Here I will only concentrate
on the case of thermal leptogenesis with $CP$ asymmetries generated
in the decays of heavy Majorana neutrinos. For a discussion on
alternative leptogenesis scenarios the reader is addressed to
Ref.~\cite{Hambye:2004fn}.}.
 The crucial ingredient in leptogenesis scenarios is the $CP$
asymmetry generated through the interference between the tree-level
and one-loop heavy Majorana neutrino decay diagrams, depicted in
Fig.~\ref{Fig1}. For the decay of the heavy Majorana neutrino $N_i$,
the $CP$ asymmetry can be expressed as~\cite{Covi:1996wh}
\begin{equation}
\label{epsi1}%
\varepsilon_i=\frac{\Gamma_i-\bar{\Gamma}_i}{\Gamma_i+\bar{\Gamma}_i}
=-\frac{3}{16\pi}\sum_{j\neq i}\frac{{\rm Im}\left[\,(Y_\nu^\dag
Y_\nu^{})_{ij}^2\,\right]}{(Y_\nu^\dag
Y_\nu)_{ii}}\,\frac{M_i}{M_j}\left(\mathcal{C}_{ij}^V+\mathcal{C}_{ij}^S\right),
\end{equation}
where $\mathcal{C}_{ij}^V$ and $\mathcal{C}_{ij}^S$ denote the
vertex and self-energy contributions, respectively. Explicitly,
these are given by:
\begin{equation}
\label{CV}%
\mathcal{C}_{ij}^S=
\frac{2}{3}\frac{M_j^2\Delta_{\!ji}}{\Delta_{ji}^2+M_i^2\Gamma_j^2}\;,\;\
\mathcal{C}_{ij}^V=\frac{2}{3}\frac{M_j^2}{M_i^2}\left[\left(1+\frac{M_j^2}
{M_i^2}\right)\log\left(1+\frac{M_i^2}{M_j^2}\right)-1\right]\,,
\end{equation}
where $\Delta_{\!ji}=M_j^2-M_i^2$ and $\Gamma_j=(Y_\nu^\dag
Y_\nu^{})_{jj}\,M_i/(8\,\pi)$ is the tree-level decay width of
$N_j$.
If the heavy Majorana neutrino masses are such that $M_1 \ll M_2,
M_3$ only the decay of the lightest Majorana neutrino $N_1$ is
relevant for the lepton asymmetry. This is a reasonable assumption
if the interactions of the lightest Majorana neutrino $N_1$ are in
thermal equilibrium at the time of the $N_{2,3}$ decays, so that the
asymmetries produced by the heaviest neutrino decays are erased
before the lightest one decays, or if $N_{2,3}$ are too heavy to be
produced after inflation \cite{Giudice:1999fb}. In this case,
$\varepsilon_1$ reduces to
\begin{equation}
\label{epsi2}%
\varepsilon_1\simeq-\frac{3}{16\pi}\sum_{j\neq 1}\frac{{\rm
Im}\left[\,(Y_\nu^\dag Y_\nu^{})_{1j}^2\,\right]}{(Y_\nu^\dag
Y_\nu)_{ii}}\,\frac{M_1}{M_j}\,.
\end{equation}
As expected, in order to produce a lepton asymmetry, $CP$ and lepton
number must be violated by the interactions of the heavy Majorana
neutrinos. The out-of-equilibrium condition for the decay of $N_1$
is attained if $\Gamma_1(M_1)<H(M_1)$, where $H(M_1)$ is the Hubble
parameter evaluated at $T=M_1$. It is instructive to define a decay
parameter $K=\Gamma_1(M_1)/H(M1)$ in such a way that the
out-of-equilibrium condition is translated into $K<1$.

The produced lepton asymmetry $Y_L$ is converted into a net baryon
asymmetry $Y_B$ through the $(B+L)$-violating sphaleron processes.
By considering the interactions in thermal equilibrium in the
thermal plasma and the correspondent balance between the chemical
potentials of the different particle species one can obtain the
following relation between the $B$, $B-L$ and $L$ number-to-entropy
density ratios~\cite{Harvey:1990qw}
\begin{align} \label{lepto6}
Y_B=a\,Y_{B-L}=\frac{a}{a-1}\,Y_L\;\;,\;\;a=
\frac{8\,N_f+4\,N_H}{22\,N_f+13\,N_H}\,,
\end{align}
where $N_f$ and $N_H$ are the number of fermion families and complex
Higgs doublets, respectively. Taking into account that $N_f=3$ and
$N_H=1$ for the SM, one gets $a \simeq 1/3$ and $Y_B \simeq
-0.5\,Y_L$. Alternatively, one can use the number-to-photon density
ratios $\eta_X=n_X/n_\gamma$ which, in the case where entropy is
conserved, are related to $Y_X$ by $\eta_X=7.04\,Y_X$. Departing
from the scenario where no asymmetry is present, the final value of
$\eta_B$ can be simply expressed by the relation: $\eta_B\simeq
-10^{-2}\,\kappa\,\varepsilon_1$, where $\kappa$ is the leptogenesis
efficiency factor. The computation of $\kappa$ requires the
numerical solution of the Boltzmann equations which take into
account the different processes in which the heavy Majorana
neutrinos are involved. The result depends on the interplay between
the terms which produce the lepton asymmetry, and those which tend
to erase it. Exhaustive discussions on these matters can be found
for instance in
Refs.~\cite{Barbieri:1999ma,Buchmuller:2002rq,Giudice:2003jh}, where
simple fits which allow to calculate $\kappa$ in some regimes are
also presented.

In a theoretical framework where neutrino masses are generated
through the seesaw mechanism, leptogenesis arises as the natural
candidate to explain the baryon asymmetry of the Universe.
Therefore, it is natural to ask whether leptogenesis is capable of
constraining the high and/or low-energy parameter spaces. In the
limit where the heavy Majorana neutrino masses are hierarchical, two
important bounds have been obtained for the light effective neutrino
masses~\cite{Buchmuller:2002rq,Giudice:2003jh,Hambye:2003rt} and for
$M_1$
~\cite{Buchmuller:2002rq,Giudice:2003jh,Hambye:2003rt,Hamaguchi:2001gw}
namely,
\begin{align}
\label{bounds}%
m_3 \lesssim 0.15~{\rm eV}\,\;,\; M_1 \gtrsim (10^8-10^9)\,{\rm
GeV}\,.
\end{align}
In particular, the upper bound on $m_i$ is of crucial importance
since future experiments will be sensitive to neutrino masses of the
order of 0.1 eV. An example of such an experiment is
KATRIN~\cite{Osipowicz:2001sq} which will probe neutrino masses down
to 0.35\,eV. Therefore, a positive signal at KATRIN would exclude
the most simple thermal leptogenesis scenario. Upcoming neutrinoless
double beta decay setups will be important as well, since they can
provide relevant information about neutrino
masses~\cite{Feruglio:2002af}.

\section{Flavor, CP violation and leptogenesis}

From the point of view of model-building, the constraints presented
above constitute a necessary condition to be fulfilled by any seesaw
model (with hierarchical heavy Majorana neutrinos) which aims at
explaining the value of the BAU. This fact has led many authors to
investigate whether there is a connection between flavor, leptonic
$CP$ violation and leptogenesis can be established or
not~\cite{Branco:2004hu}. A very simple example of such interplay
has been presented in Ref.~\cite{Branco:2002kt} where the flavor
structure of the high-energy neutrino sector is similar to the quark
one. The Dirac neutrino Yukawa coupling matrix reads:
\begin{align} \label{gf13}
v\,Y_\nu=V_L^{\dag}\,d\,U_R\,,\,d =\text{diag} (m_u,m_c,m_t)\,,
\end{align}
where $m_{u,c,t}$ are the up, charm and top quark masses. The matrix
$V_L$ is analogous to the CKM matrix of the quark sector. In the
case where the small mixing in $V_L$ is neglected, the heavy
Majorana masses are approximately given by
\begin{align} \label{an14}
M_1 \simeq \frac{m_u^2}{s_{12}^2\sqrt{\dmtwon}+ \sqrt{\dmthtw}\,
s_{13}^2}\simeq 3.3\times 10^5\left(\frac{m_u}{1\,{\rm
MeV}}\right)^2\,{\rm GeV}
\end{align}
for a hierarchical spectrum of the light neutrinos $(m_1\rightarrow
0)$. It is clear that, for an up-quark mass $m_u(M_1)\simeq 1~{\rm
MeV}$, this value is far below the lower bound given in
(\ref{bounds}) for $M_1$. In this case, $\eta_B$ has the maximum
value:
\begin{align} \label{an15}
\eta_B^{\rm max} \simeq 9.5\times 10^{-13}
\left(\frac{m_u}{1\,{\rm MeV}}\right)^2\,,
\end{align}
which is three orders of magnitude below the experimental value
shown in (\ref{BAU}). For an inverted-hierarchical neutrino mass
spectrum, Eq.~(\ref{an14}) remains valid, but the maximum value of
$\eta_B$ is decreases by two orders of magnitude. A similar
situation occurs in a model based in the assumption of democratic
flavor structures in the leptonic sector~\cite{Akhmedov:2000yt}. The
above discussion shows that the hierarchy in the Dirac neutrino
Yukawa couplings has to be weaker than the one observed in the quark
sector in order for leptogenesis to be successful\footnote{Some
exceptions can be found if one considers special relations between
some elements of the effective neutrino mass
matrix~\cite{Akhmedov:2003dg}. In this case, $M_1 \simeq M_2$,
leading to an enhancement of the $CP$ asymmetries in the decays of
$N_1$ and $N_2$.}.

Regarding the connection between the $CP$ violation needed for
leptogenesis and that potentially measured in neutrino experiments
one can see that, while leptogenesis only depends on the phases of
the matrix $V_L$ shown in Eq.~(\ref{gf13}), the low-energy phases
$\delta$, $\alpha$ and $\beta$ of Eq.~(\ref{Uparame}) depend on both
$V_L$ and $U_R$. Using the Casas-Ibarra parametrization defined in
Eq.~(\ref{ynuR}) one can show that only the phases of $\mathcal{O}$
are relevant for leptogenesis since
\begin{align}
{\rm Im}[(Y_\nu^\dag
Y_\nu^{})_{ij}^2]=\frac{M_i\,M_j}{v^4}\sum_{a,b}m_a m_b\,{\rm
Im}\left(\mathcal{O}_{ia}^\dag\mathcal{O}_{aj}\mathcal{O}_{ib}^\dag
\mathcal{O}_{bj}\right)\,.
\end{align}
In particular, this means that one may have viable leptogenesis even
in the limit where there are no $CP$-violating phases (neither Dirac
nor Majorana) in the PMNS mixing matrix $U$ and hence, no $CP$
violation at low energies \cite{Rebelo:2002wj}. Explicitly defining
$\mathcal{O}\equiv|\mathcal{O}_{ij}| e^{i \varphi_{ij}/2}$, the
$CP$-asymmetry $\varepsilon_1$ reads~\cite{Branco:2002xf}
\begin{align}
\varepsilon_1 \simeq \frac{3}{16\pi} \frac{M_1}{v^2} \frac { \dmthtw
|\mathcal{O}_{31}|^2 \sin \varphi_{31}  - \dmtwon
|\mathcal{O}_{11}|^2 \sin\varphi_{11}} { m_1|\mathcal{O}_{11}|^2
+m_2|\mathcal{O}_{21}|^2 +m_3|\mathcal{O}_{31}|^2}\,, \label{edb}
\end{align}
where $\dmtwon\simeq 8\times 10^{-5}~{\rm eV}^2$ and $\dmthtw\simeq
2\times 10^{-3}~{\rm eV}^2$~\cite{Strumia:2005tc} are the neutrino
mass squared differences measured in neutrino oscillation
experiments. The above equation recovers what one would have
expected by intuition, namely that the physical quantities involved
in determining $\varepsilon_1$ are just $M_1$, the spectrum of the
light neutrinos, $m_i$, and the first column of $\mathcal{O}$, which
expresses the composition of the lightest heavy Majorana neutrino in
terms of the light neutrino masses $m_i$. Therefore, in general it
is not possible to establish a link between low-energy CP violation
and leptogenesis. This connection is model dependent: it can be
drawn only by specifying a particular \emph{ansatz} for the
fundamental parameters of the seesaw.

In terms of $CP$-violating invariants, it has been shown
\cite{Branco:2001pq} that the strength of CP violation at low
energies, observable for example through neutrino oscillations, can
be obtained from the following low-energy weak-basis (WB) invariant:
\begin{align}
\mathcal{ T}_{CP} = {\rm Tr}\left[\,\mathcal{H}, H_\ell
\,\right]^3=6\,i \,\Delta_{21}\,\Delta_{32}\,\Delta_{31}\,{\rm Im}
\left[\, \mathcal{H}_{12}\,\mathcal{H}_{23}\,\mathcal{H}_{31}\,
\right]\,, \label{TCP}
\end{align}
where $\mathcal{H}=\mathcal{M}\,\mathcal{M}^{\dag}$,
$H_\ell=Y_\ell\,{Y_\ell}^{\dagger}$ and
$\Delta_{21}=({y_{\mu}}^2-{y_e}^2)$ with analogous expressions for
$\Delta_{31}$, $\Delta_{32}$. This relation can be computed in any
weak basis. The low-energy invariant in Eq.~(\ref{TCP}) is sensitive
to the Dirac-type phase $\delta$ and vanishes for $\delta=0$. On the
other hand, it does not depend on the Majorana phases $\alpha$ and
$\beta$ appearing in the leptonic mixing matrix. The quantity ${\cal
T}_{CP}$ can be fully written in terms of physical observables once
\begin{align}
{\rm Im} \left[\, \mathcal{H}_{12}\,\mathcal{H}_{23}
\,\mathcal{H}_{31}\, \right] = -
\dmtwon\,\dmthon\,\dmthtw\,\mathcal{J}_{CP}\,, \label{ImHHH}
\end{align}
where the $\Delta m_{ij}^2$'s are the usual light neutrino mass
squared differences and $\mathcal{ J}_{CP}$ is the imaginary part of
an invariant quartet of $U$ appearing in the difference of the
$CP$-conjugated neutrino oscillation probabilities
$P(\nu_e\rightarrow\nu_\mu)-P(\bar{\nu}_e\rightarrow
\bar{\nu}_\mu)$. One can conveniently write~\cite{Branco:2002xf}
\begin{align} \mathcal{J}_{CP}=-\frac{{\rm Im} \left[\,\mathcal{H}_{12}
\,\mathcal{H}_{23}\,\mathcal{H}_{31}\,
\right]}{\dmtwon\,\dmthon\,\dmthtw}\,, \label{Jfin}
\end{align}
which allows the computation of the low-energy CP invariant without
resorting to the mixing matrix $U$.

It is also possible to write WB invariants which are particularly
useful to leptogenesis \cite{Branco:2001pq}. The requirement of $CP$
invariance implies the vanishing of the following WB invariants
\begin{align}
\label{II.1:I1}%
I_1&\equiv{\rm
Im}\left[\,\Tr\left(H_\nu\,H_R\,M_R^\ast\,H_\nu^\ast\,M_R
\,\right)\,\right]\,,\nonumber\\I_2&\equiv {\rm
Im}\left[\,\Tr\left(H_\nu\,H_R^2\,M_R^\ast\,H_\nu^\ast\,M_R
\,\right)\,\right]\,,\nonumber\\I_3&\equiv {\rm
Im}\left[\,\Tr\left(H_\nu\,H_R^2\,M_R^\ast\,H_\nu^\ast\,M_R\,H_R\,\right)
\,\right]\,,
\end{align}
where $H_\nu \equiv Y_\nu^\dag Y_\nu^{}\;,\;H_R \equiv M_R^\dag
M_R^{}$. Unless any of the $M_i$ vanish or one is in the case of
right-handed neutrino degeneracy, the conditions $I_k=0$ require the
trivial solution ${\rm Im}[(H_\nu)_{ij}^2]=0$. In terms of $I_1$,
$I_2$ and $I_3$ one has
\begin{align}
\label{II.2:Hij}%
&{\rm
Im}[(H_\nu)_{12}^2]=\frac{I_3-I_2M_3^2+I_1M_3^4}{M_1\,M_2\, \Delta_{21}
\, \Delta_{31}\,\Delta_{32}}\,,\nonumber\\
-&{\rm Im}[(H_\nu)_{13}^2]=\frac{I_3-I_2M_2^2+I_1M_2^4}{M_1\,M_3\,
\Delta_{21}\,\Delta_{31}\,\Delta_{32}}\,,\nonumber\\
&{\rm
Im}[(H_\nu)_{23}^2]=\frac{I_3-I_2M_1^2+I_1M_1^4}{M_2\,M_3\,\Delta_{21}
\,\Delta_{31}\,\Delta_{32}}\,.
\end{align}
In the limit of hierarchical $M_i$, the $CP$-asymmetry
$\varepsilon_1$ can be expressed as a function of the three
invariants $I_i$ and of the masses $M_i$
\begin{align}
\label{II.2:eps1inv}%
\varepsilon_1&\simeq
-\frac{3}{16\,\pi\,(H_\nu)_{11}}\frac{I_3+I_2\,M_3^2+I_1\,M_3^4}
{M_2^4\,M_3^4}\,.
\end{align}

A very simple reasoning shows that a general relation between lepton
flavor violation in rare decays of the type $\ell_i\rightarrow
\ell_j\,\gamma$ and leptogenesis does not exist. Taking as an
example the minimal supergravity case, the branching ratios of these
processes can be approximated by~\cite{Casas:2001sr}:
\begin{align}
\label{LFVap}%
{\rm BR}(\ell_i\rightarrow \ell_j\,\gamma)\simeq
\frac{\alpha^3}{G_F^2
m_S^8}\,\left|\frac{3\,m_0^2+A_0^2}{8\pi^2}\,(Y_\nu^{}Y_\nu^\dag)_{ij}
\log\left(\frac{M_X}{M}\right)\right|^2\tan^2\beta\,,
\end{align}
where $m_0$ and $A_0$ are the universal soft mass and trilinear
coupling at a high scale $M_X$, and $m_S$ is an average slepton
mass. For simplicity I have taken the case of degenerate heavy
Majorana neutrinos. From the above equation, it is straightforward
to see that the flavor violation effects induced in the slepton mass
matrix are sensitive to the left-handed rotation $V_L$ given in
Eq.~(\ref{gf13}). Thus, leptogenesis depends on $U_R$, one
immediately concludes that a general connection between the LFV
decay rates and the value of the BAU cannot be established. This
fact is illustrated by the numerical computations presented in
Refs.~\cite{Ellis:2002xg} for both the cases of hierarchical and
quasi-degenerate heavy Majorana neutrino mass spectra. Nevertheless,
in some interesting cases some relation between both phenomena may
be observed~\cite{Pascoli:2003uh}.

\section{Radiative leptogenesis}
In supersymmetric theories, the constraints on $M_1$ shown in
(\ref{bounds}) may be in conflict with the upper bound on the
reheating temperature of the Universe, which can be as low as
$10^6$~GeV~\cite{Kawasaki:2004yh}. One of the possible solutions to
this problem is to consider the case of quasi-degenerate heavy
Majorana neutrinos, which can be perfectly reconciled with the
available neutrino data~\cite{GonzalezFelipe:2001kr}. In such
scenarios, leptogenesis may work with heavy neutrino masses as low
as 1~TeV, due to an enhancement of the $CP$-asymmetries generated in
the $N_i$-decays~\cite{Pilaftsis:2003gt}.

A natural way to generate the small heavy Majorana mass splittings
is through renormalisation group
effects~\cite{GonzalezFelipe:2003fi}. In order to illustrate how
this mechanism works, let me consider the case of two $N_i$. At a
scale $\Lambda_D$ the heavy neutrinos are degenerate, \emph{i.e.}
$M_1=M_2 \equiv M$, with $M < \Lambda_D$. In this limit, $CP$ is
not necessarily conserved. This is supported from the fact that
the weak-basis invariant $\mathcal{J}_1=M^{-6}{\rm Tr}\left[ Y_\nu
Y_\nu^{T} Y_{\ell}^{}Y_{\ell}^{\dagger} Y_\nu^{*}
Y_\nu^{\dagger},Y_{\ell}^{*}Y_{\ell}^{T}\right] ^{3}$, which is
not proportional to $M_{2}^2-M_{1}^2$, does not vanish in the
exact degeneracy limit. On the other hand, a non-zero leptonic
asymmetry can be generated if and only if the $CP$-odd invariant
\begin{equation}
\mathcal{J}_{2}={\rm Im\,Tr}\,[H_\nu M_R^{\dagger} M_R
M_R^{\dagger}H_\nu^T M_R ]=M_1 M_2 (M_2^2-M_1^2)\, {\rm
Im\,}[(H_\nu)_{12}^2]\,,
\end{equation}
does not vanish~\cite{Pilaftsis:2003gt}. The condition
$\mathcal{J}_{2}\neq 0$ requires both $M_1 \neq M_2$ and ${\rm
Im\,}[(H_\nu)_{12}^2] \neq 0$, at the leptogenesis scale $M$. These
requirements are guaranteed by the running of $M_R$ and $Y_\nu$ from
$\Lambda_D$ to $M$. The renormalisation group equations (RGE) for
$Y_\nu$, $H_\nu$ and the heavy neutrino masses $M_i$
are~\cite{Casas:1999tp}:
\begin{align}
\label{RGEMRdiag1} \frac{d Y_\nu}{dt}&=k\,Y_\nu+\left[ -a\,Y_\ell^{}
Y_\ell^\dagger-b\,Y_\nu^{} Y_\nu^\dagger \right] Y_\nu + Y_\nu T \,,
\\
\label{RGEMRdiag2} \frac{d H_\nu}{dt}&= 2 \,k\,H_\nu-
2\,b\,H_\nu^2-2a Y_\nu^\dag Y_\ell^{} Y_\ell^\dag Y_\nu
\!+[H_\nu,T]\,,\\
\label{RGEMRdiag} \frac{d M_i}{dt}&=2\,
c\,M_i\,,(H_\nu)_{ii}\;,\;[H_\nu,T]=H_\nu T-TH_\nu\,,
\end{align}
where $k$ is a function of ${\rm Tr}(Y_X^{}Y_X^\dag)$ and the
gauge couplings~\cite{Chankowski:2001mx}. The factors $a$, $b$ and
$c$ are $a_{{\rm SM}}=-b_{{\rm SM}}=3/2\;,\;b_{{\rm
MSSM}}=3\,a_{{\rm MSSM}}=-3\;,\;c_{\rm MSSM}=2\,c_{\rm SM}=2$ for
the SM and MSSM cases. The anti-Hermitian matrix $T$ encodes the
effects of rotation to the basis where $M_R$ is diagonal. Defining
the degree of degeneracy between $M_1$ and $M_2$ through the
parameter $\delta_N \equiv M_2/M_1-1$ one has:
\begin{equation} \label{Rmatrix}
T_{12}=\frac{2+\delta_N}{\delta_N}\, {\rm Re}\,[H_{12}]+
i\frac{\delta_N}{2+\delta_N}\, {\rm Im}\,[H_{12}]\;,\;T_{ii}=0\,.
\end{equation}
From Eq.~(\ref{Rmatrix}) on can see that if $\delta_N=0$ at a
given scale $\Lambda_D$, then the RGE in Eqs.~(\ref{RGEMRdiag1})
and (\ref{RGEMRdiag2}) become singular, unless one imposes ${\rm
Re}\,(H_{12})=0$. This can be achieved by rotating the heavy
fields by an orthogonal transformation $O$, being the rotation
angle $\theta$ such that $\tan 2\theta = 2\,{\rm
Re}\,[H_{12}]/(H_{22}-H_{11})$. Under this transformation, $Y_\nu
\rightarrow Y_\nu^\prime=YO$ and $H_\nu \rightarrow
H_\nu^\prime={Y_\nu^\prime}^\dag Y_\nu^\prime=O^\dagger H_\nu O$.
The scale dependence of the degeneracy parameter $\delta_N$ is
governed by
\begin{equation}
\frac{d
\delta_N}{dt}=
2\,c\,(\delta_N+1)[(H_\nu)_{22}-(H_\nu)_{11}+2\Delta]\,,
\end{equation}
with $\Delta \equiv \tan \theta\, {\rm Re}\,[H_{12}]$. In the limit
$\delta_N \ll 1$, the leading-log approximation for $\delta_N(t)$
can be easily found to be
\begin{equation}
\delta_N(t)\simeq 2\,c\,[(H_\nu)_{22}-(H_\nu)_{11}+2\Delta]\,t\,.
\end{equation}

For quasi-degenerate Majorana neutrinos the $CP$-asymmetries
generated in their decays are approximately given
by~\cite{Pilaftsis:2003gt}
\begin{equation}\label{e12d2}
\varepsilon_j\simeq\frac{{\rm
Im}\,[\,H^{\prime\,2}_{21}]}{16\,\pi\,\delta_N\,H^{\prime}_{jj}}\!\!\left(1
+
\frac{\Gamma_i^2}{4M^2\delta_N^2}\right)^{-1}\!\!\!\!\,,i,j=1,2\,(i\neq
j)\,,
\end{equation}
where the $\Gamma_{i}$ are the heavy Majorana decay widths
introduced in the previous section. The above equation shows that
\begin{equation}
\varepsilon_i(t) \propto {\rm
Im}\,[(H_\nu^{\prime})_{12}\,(t)\,]\,{\rm
Re}\,[(H_\nu^{\prime})_{12}\,(t)\,]\,\,,\quad i=1,2\,.
\end{equation}
Therefore, a necessary condition to have a nonzero $CP$-asymmetry at
a given $t$ is that ${\rm Re}\,[(H_\nu^{\prime})_{12}\,(t)\,] \neq
0$. Since ${\rm Re}\,[(H_\nu^{\prime})_{12}\,(0)] = 0$, one has to
rely on running effects to generate a nonzero ${\rm
Re}\,[(H_\nu^{\prime})_{12}\,]$. From Eqs.~(\ref{RGEMRdiag2}) and
(\ref{Rmatrix}) and taking into account that ${\rm
Re}\,[(H_\nu^{\prime})_{12}\,(0)] = 0$,
\begin{equation}
\label{rehnu} {\rm Re}[(H_\nu^{\prime})_{12}(t)] \simeq
-\frac{a\,y_\tau^2}{16\pi^2}\,{\rm Re}\,[Y_{31}^{\prime\ast}
\,Y_{32}^\prime]\,t\,.
\end{equation}
The radiatively generated $\varepsilon_{1,2}$ can be computed from
Eqs.~(\ref{e12d2}) and (\ref{rehnu}).

\begin{figure*}\begin{center}
\begin{tabular}{ccc}
\hspace{-0.5cm}\includegraphics[width=5.9cm]{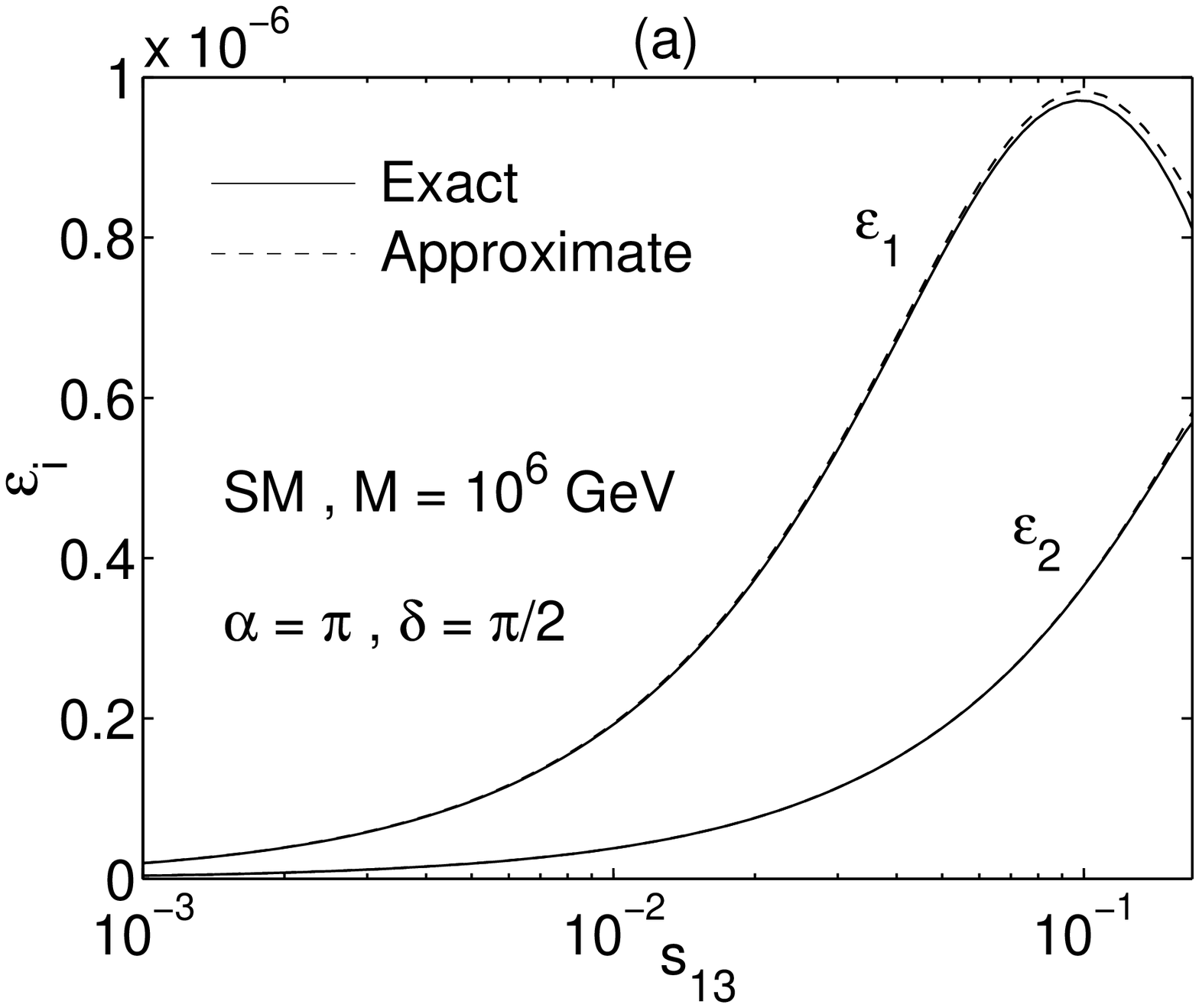}
&\includegraphics[width=5.9cm]{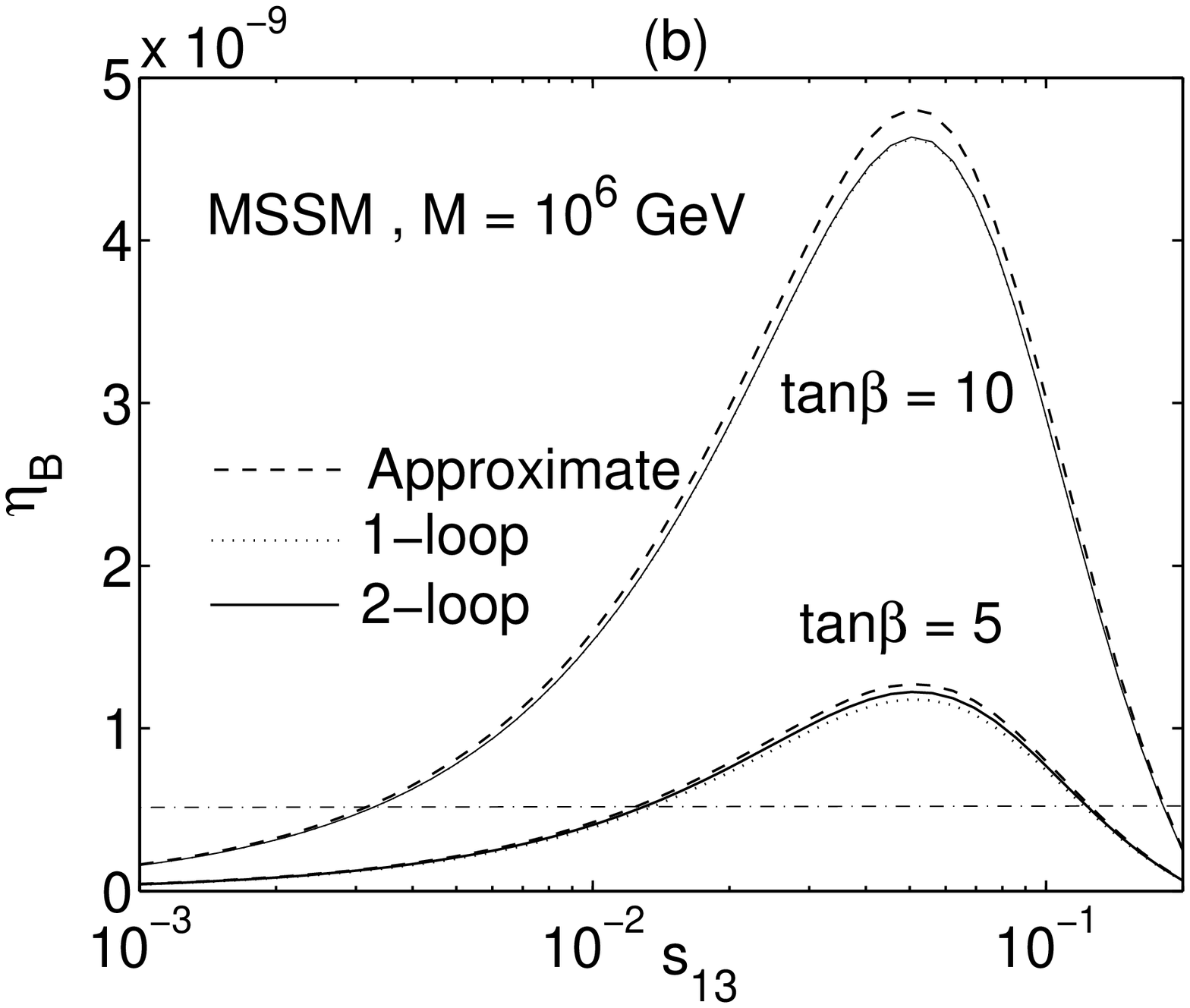}
\end{tabular}\end{center}
\caption{(a) Maximum $CP$ asymmetries $\varepsilon_{1,2}$ as a
function of $s_{13}$ in the SM case. (b) Baryon asymmetry as a
function of $s_{13}$ in the MSSM case for $\tan\beta=5,10$. The
dotted (solid) line refers to the result using the one-loop
(two-loop) RGE while the dashed line corresponds to the case where
the analytical expressions for the $CP$-asymmetries are used. The
horizontal line indicates the mean experimental value for $\eta_B$.}
\label{Fig2}\end{figure*}

In the following I will illustrate how the mechanism described above
works for a specific example. It is convenient to define the
$3\times3$ seesaw operator $\kappa$ at $\Lambda_D\,$,
$\kappa=Y_\nu^{} \,Y_\nu^T/M$, where $(Y_\nu)_{ij}=y_0\,y_{ij}$ is a
$3\times 2$ complex matrix. In order to reconstruct the high energy
neutrino sector in terms of the low energy parameters, I choose
$y_{12}=0$. The effective neutrino mass matrix $\mathcal{M}$ is
\begin{equation}
\label{Mnu1} \mathcal{M}=m_3\,U{\rm
diag}(0,\rho\,e^{i\alpha},1)\,U^T\;,\;\rho \equiv
m_2/m_3\;,\;\rho=\sqrt{\frac{\dmthon}{\dmthtw}}
\end{equation}
where $m_3$ is the mass of the heaviest neutrino and  $\alpha$ is
a Majorana phase. In the present case, the maximum $CP$-asymmetry
$\varepsilon_1$ is approximately given by
\begin{equation}
\label{e1max}
    \varepsilon_1^{{\rm max}} \simeq -\frac{3y_\tau^2\,c_{12}}{128\pi} \frac{
(1+\rho)}{(1-\rho)} \simeq -10^{-6}\,,
\end{equation}
which is in perfect agreement with the result shown in
Fig.~\ref{Fig2}.(a). After taking into account the washout effects,
the final value of the baryon asymmetry is $\eta_B^{\text{max}}
\simeq 3 \times 10^{-10}$, which is smaller than the experimental
result by a factor of two\footnote{The SM result can be reconciled
with the experimental by introducing a third heavy Majorana neutrino
much heavier than $N_{1,2}$~\cite{Branco:2005ye}}.
In Fig.~\ref{Fig2}.b $\eta_B$ is computed for the MSSM case. Besides
the factor of two which has to be included in Eq.~(\ref{e12d2}) due
to the presence of supersymmetric particles in the decays, weone
expects an extra enhancement factor of $(1+\tan^2\beta)$ in the MSSM
respective to the SM case since $\varepsilon_{1,2} \propto
y_\tau^2$~(see also Ref.~\cite{Turzynski:2004xy}). It can be seen
that, depending on the value of $\tan\beta$, the maximum of $\eta_B$
can be far above the experimental value. An interesting feature of
radiative leptogenesis is that the BAU is practically independent
from the gap between $\Lambda_D$ and $M$ which means that $M$ can be
as low as $1$~TeV.

\section{Concluding remarks}

The idea that the origin of the matter-antimatter asymmetry of the
Universe may be related with the mechanism through which neutrinos
become massive has led to an intense investigation in the last few
years. Although it will be difficult to establish leptogenesis as
being responsible for the generation of the BAU, future neutrino
experiments will be able to rule out its simplest variant. This
would be the case if the absolute neutrino mass scale happens to be
above $\sim 0.2$~eV. On the other hand, positive signals of heavy
Majorana neutrinos in future colliders~\cite{delAguila:2005pf} could
open the window to low-scale leptogenesis
mechanisms~\cite{Pilaftsis:2005rv}.

Upcoming neutrino oscillation experiments will, under certain
conditions, be sensitive to $CP$-violating effects in the leptonic
sector. Moreover, an experimental indication in favor of
neutrinoless double beta decays processes would reveal the Majorana
nature of neutrinos~\cite{Schechter:1981bd} and possibly give some
information about the Majorana phases~\cite{Pascoli:2005zb}. Even in
the most favorable scenario where the strength of leptonic $CP$
violation is measured, it seems to be difficult to conclude whether
this $CP$-violating effects are relevant for leptogenesis or not.
This stems from the fact that high and low-energy neutrino
parameters are not connected in a model-independent way. Obviously,
such statements are based in our present understanding of these
phenomena. Hopefully, future data from several particle and
cosmological experiments, as well as novel theoretical approaches,
may reveal unequivocal signals confirming leptogenesis as the
mechanism responsible for the generation of the BAU.

\vspace{0.3cm} \noindent {\bf Acknowledgements}\vspace{0.3cm}

I am grateful to the Organisers of this Symposium for the
invitation to contribute to these Proceedings and to all my
collaborators with whom part of the work presented here has been
done. In this special occasion, I devote a special word of
gratitude to Gustavo Branco wishing him many more successes for
the future.

This work was supported by {\it Funda\c c\~ao para a Ci\^encia e
Tecnologia} under the grant SFRH/BPD/14473/2003.

\end{document}